\documentclass[%
 reprint,
 amsmath,amssymb,
 aps,
 pra,
]{revtex4-1}

\usepackage{graphicx}
\usepackage{dcolumn}
\usepackage{bm}
\usepackage{multirow}
\usepackage{siunitx}
\usepackage{epstopdf}
\usepackage{color}

\begin{document}


\title{A Diamond-Photonics Platform Based on Silicon-Vacancy Centers in a Single Crystal Diamond Membrane and a Fiber-Cavity}

\author{
Stefan H\"au\ss{}ler$^{1,2}$,
Julia Benedikter$^{3,4}$,
Kerem Bray$^{5}$,
Blake Regan$^{5}$,
Andreas Dietrich$^{1}$,
Jason Twamley$^{6}$,
Igor Aharonovich$^{5}$,
David Hunger$^{7}$ and
Alexander Kubanek$^{1,2}$
}

\email[]{alexander.kubanek@uni-ulm.de}

\affiliation{$^1$ Institute for Quantum Optics, Ulm University, Albert-Einstein-Allee 11, D-89081 Ulm, Germany}
\affiliation{$^2$ Center for Integrated Quantum Science and Technology (IQst), Ulm University, Albert-Einstein-Allee 11, D-89081 Ulm, Germany}
\affiliation{$^3$ Fakult\"at f\"ur Physik, Ludwig-Maximilians-Universit\"at, Schellingstra\ss{}e 4, D-80799 M\"unchen, Germany}
\affiliation{$^4$ Max-Planck-Institut f\"ur Quantenoptik, Hans-Kopfermann-Stra\ss{}e 1, D-85748 Garching, Germany}
\affiliation{$^5$ School of Mathematical and Physical Sciences, University of Technology Sydney, Ultimo, New South Wales 2007, Australia}
\affiliation{$^6$ Centre for Engineered Quantum Systems, Department of Physics and Astronomy, Macquarie University, Sydney, New South Wales 2109, Australia}
\affiliation{$^7$ Physikalisches Institut, Karlsruher Institut f\"ur Technologie, Wolfgang-Gaede-Stra\ss{}e 1, D-76131 Karlsruhe, Germany}

\date{\today}

\begin{abstract}

We realize a potential platform for an efficient spin-photon interface, namely negatively-charged silicon-vacancy centers in a diamond membrane coupled to the mode of a fully-tunable, fiber-based, optical resonator. We demonstrate that introducing the thin ($\sim 200 \, \text{nm}$), single crystal diamond membrane into the mode of the resonator does not change the cavity properties, which is one of the crucial points for an efficient spin-photon interface. In particular, we observe constantly high Finesse values of up to $3000$ and a linear dispersion in the presence of the membrane. We observe cavity-coupled fluorescence from an ensemble of SiV$^{-}$ centers with an enhancement factor of $\sim 1.9$. Furthermore from our investigations we extract the ensemble absorption and extrapolate an absorption cross section of $(2.9 \, \pm \, 2) \, \cdot \, 10^{-12} \, \text{cm}^{2}$ for a single SiV$^{-}$ center, much higher than previously reported.

\end{abstract}

\maketitle


\section{Introduction}

Color centers in diamond are amongst the most promising candidates for the realization of quantum repeaters that are needed for the distribution of quantum states over long distances and secure quantum communication using quantum key distribution \cite{briegel1998quantum, takeoka2014fundamental, luong2016overcoming}. So far, major efforts were dedicated to NV$^{-}$ centers, resulting in the recent demonstration of deterministic remote entanglement using diamond spin qubits, which states a key building block for an extended quantum network \cite{humphreys2018deterministic}. A complementary platform can be realized utilizing the negatively-charged silicon-vacancy center (SiV$^{-}$). Its atom-like optical properties, such as large Debye-Waller factor, spectral stability and a narrow inhomogeneous distribution, have recently attracted great attention \cite{rogers2014multiple, sipahigil2014indistinguishable, hepp2014electronic, schroder2017scalable, zhang2017complete}. Also, long spin life- and coherence times of $T_{1} = 1 \, \text{s}$ and $T_{2} = 13 \, \text{ms}$ at temperatures of $100 \, \text{mK}$ were demonstrated, completing all requirements needed to realize a quantum repeater \cite{Sukachev2017silicon, Becker2018alloptical}. A bottleneck for a scalable implementation remains the low extraction efficiency of coherent photons. The efficiency problem can be counteracted by Purcell-enhanced photon emission when coupling the SiV$^{-}$ center to the mode of an optical resonator \cite{sipahigil2016anintegrated}. Fiber-based Fabry-Perot resonators are particularly suitable, since they offer full frequency tuneability and single photon emission directly into the fiber mode. In pioneering work, single SiV$^{-}$ centers in nanodiamonds were coupled to the mode of a fiber-based resonator \cite{benedikter2017cavity}. However, light scattering restricts the Finesse of the coupled system. Diamond membranes hosting nitrogen vacancy centers have been integrated into the mode of an optical resonator leading to a modified dispersion relation due the composite diamond-air cavity system \cite{janitz2015fabry, riedel2017deterministic, bogdanovic2017design} and restricting the length of the resonator to the optical thickness of the membrane, the ablation depth and the electric field penetration into the mirror coatings. \\
\begin{figure*}[htbp]
	\includegraphics[width=0.92\textwidth]{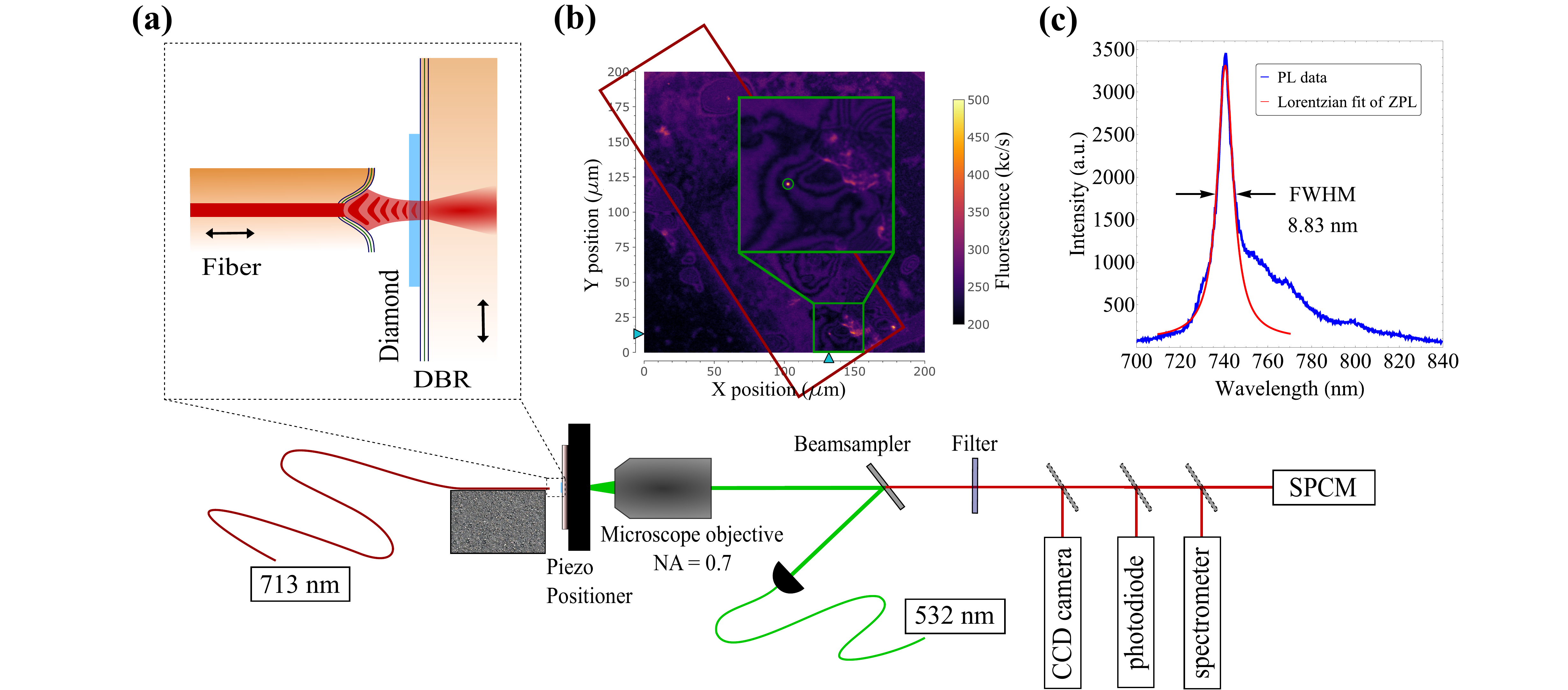}
		\caption{(a) Schematic sketch of the setup for the cavity measurements. The fiber mirror is mounted on a large damping block to reach high passive stability. We use a piezoelectric shear actuator to tune the cavity length. The flat mirror with the thin diamond membrane is mounted on a two-axis piezo positioning stage to allow lateral positioning. We use a $\text{NA} = 0.7$ microscope air objective to excite color centers through the flat mirror and collect the outcoupled cavity mode. We analyse the cavity properties by measuring transmission of a $713 \, \text{nm}$ laser with a photodiode and CCD camera. Flourescence is measured with a single photon counting module (SPCM) and spectrometer. (b) Confocal image with red rectangle indicating the diamond membrane. The position of the SiV$^{-}$ ensemble is marked as green circle in the inset. (c) Photoluminescence (PL) spectrum of the SiV$^{-}$ ensemble. A Lorentzian fit of the ZPL at $738 \, \text{nm}$ yielding a FWHM linewidth of $8.83 \, \text{nm}$.}
	\label{fig:confocal}
\end{figure*}
Here, we investigate a novel spin-photon interface based on the SiV$^{-}$ center in a thin single crystal diamond membrane with a thickness of approximately $200 \, \text{nm}$. We show that our platform gives access to smallest resonator length with the dispersion relation of an empty resonator and with minimal scattering loss. From the coupled SiV-cavity system we extract the absorption cross section of SiV$^{-}$ to $(2.9 \, \pm \, 2) \, \cdot \, 10^{-12} \, \text{cm}^{2}$.

\section{Methods}
 
Our experiments are performed using a tunable fiber-based Fabry-Perot cavity. The cavity system consists of a macroscopic flat mirror substrate on the one side and the tip of a multimode optical fiber on the other side. The fiber tip is provided with a concave structure, which is fabricated using laser ablation of a high power $\text{CO}_{2}$ laser. The ablation process exploits the high absorption coefficient of the fiber material in the mid infrared wavelength regime leading to melting and evaporation of the fiber tip. As a result a Gaussian-shaped profile with extremely low surface roughness is produced. The radius of curvature (ROC) of the structure is estimated from a parabolic fit to an interferometer image and is chosen rather large ($ROC = 90 \, \mu\text{m}$) to allow large cavity lengths of up to $\approx 65 \, \mu\text{m}$ in stable operation. The fiber tip is coated with a dielectric mirror stack (DBR) with high reflectivity between $690 \, \text{nm}$ and $890 \, \text{nm}$ with a transmission of $\approx 500 \, \text{ppm}$ at $710 \, \text{nm}$ and $\approx 30 \, \text{ppm}$ at $740 \, \text{nm}$. The flat mirror is provided with a high index terminated DBR stack to reach high reflectivity between $560 \, \text{nm}$ and $760 \, \text{nm}$. The transmission is $\approx 500 \, \text{ppm}$ at $710 \, \text{nm}$ and $\approx 3000 \, \text{ppm}$ at $740 \, \text{nm}$. The asymmetry of the mirror coatings leads to a maximum transmission at a wavelength of $710 - 715 \, \text{nm}$ and predominant emission of the coupled SiV$^{-}$ color center through the flat mirror side. Both dielectric coatings are transparent for the green light of our $532 \, \text{nm}$ laser, enabling excitation of the diamond membrane from two sides. In our case, both excitation and collection of the cavity mode is done with a $\text{NA} = 0.7$ microscope air objective via the flat mirror side. \\
The single crystal diamond membranes were created from a bulk type IIa CVD diamond from Element $6^{\text{TM}}$, with a nitrogen concentration $< 1 \, \text{ppm}$. The crystal was implanted with $1 \, \text{MeV}$ He$^{+}$ ions ($5 \cdot 10^{16} \, \text{ions/cm}^{2}$). The implanted diamond was subsequently annealed at $900 \, ^{\circ}\text{C}$ in vacuum to create a thin amorphous carbon layer $1.7 \, \mu\text{m}$ below the surface, which enables lift-off. To selectively remove the carbon layer, electrochemical etching was carried out by immersing the sample in deionized water and applying a constant forward bias of $60 \, \text{V}$. The membrane was then epitaxially overgrown using a Microwave Plasma Chemical Vapor Deposition (MPCVD), flipped and thinned to a final thickness of $\sim 200 \, \text{nm}$. SiV color centers are incorporated into the growing membrane by diffusion from the residual silicon present in the chamber. The membranes were transferred using a liquid droplet to the flat mirror. More information on the sample preparation is given in the Supplement Material and in the references \cite{magyar2011fabrication, lee2012coupling, aharonovich2012homoepitaxial, bray2018single}. \\
We use a multi-axis piezo positioner with sub-nanometer resolution to laterally place the diamond membrane in the cavity mode and a shear piezo actuator to tune the cavity length. The fiber mirror is mounted on a damping block to achieve high passive stability. The cavity transmission is characterized using a CCD camera and an avalanche photodiode while the fluorescence of the diamond membrane is measured with a single photon counting module (SPCM) and a grating spectrometer. A schematic sketch of the setup is shown in figure \ref{fig:confocal}(a). \\
We first orient the sample so that the diamond membrane faces the microscope objective and study the properties of the thin diamond membrane in confocal microscopy. A detailed confocal map of $\sim 2 \, \text{mm}^{2}$ size around the diamond membrane is recorded and used later to localize the SiV$^{-}$ ensemble in the cavity mode. A section of the $\sim 200 \times 100 \, \mu\text{m}^{2}$ diamond membrane can be seen in figure \ref{fig:confocal}(b). Figure \ref{fig:confocal}(c) shows the photoluminescence (PL) spectrum of an ensemble of SiV$^{-}$ centers inside the diamond membrane. The lateral position of the spot is marked in the inset of figure \ref{fig:confocal}(b). The ensemble zero-phonon line (ZPL) is at $738 \, \text{nm}$ with a FWHM linewidth extracted from a Lorentzian fit of $8.83 \, \pm \, 0.13 \, \text{nm}$. The phonon sideband (PSB) extends to $\sim 850 \, \text{nm}$. A fluorescence background originating from other color centers (for example NV centers) is not visible.

\section{Characterization of the diamond-membrane-cavity-system}

After flipping the flat sample mirror and closing the resonator we characterize the cavity-membrane dispersion relation by determining the fundamental mode frequencies of the cavity in transmission via a white Light Emitting Diode (LED) source. The transmitted light is collected with the microscope objective on the flat mirror side and analyzed with the spectrometer using a $1200 \, \text{groves/mm}$ grating. We observe several longitudinal resonances when the cavity length is changed in $\approx 20 \, \text{nm}$ steps while continuously measuring the transmitted light with the spectrometer. For a bare cavity without membrane the resonance frequency $\nu_{m}$ of the fundamental mode $m$, including the Gouy phase shift, is  
\begin{equation}
\nu_{m} = \frac{c}{2L} \left(m + \frac{\cos^{-1}(\sqrt{1-L/ROC})}{\pi} \right),
\end{equation}
where $L$ is the effective length of the cavity, including the field penetration into the coating of the DBR. Additionally, we observe higher order transverse modes with distinct eigenfrequencies. \\
Inserting a diamond sample into the cavity typically changes the characteristics of the resonance frequencies. The behaviour can be described quantitatively by a one-dimensional model (see \cite{janitz2015fabry}). Considering a diamond membrane of thickness $t_{d}$, and refractive index $n_{d}$, and an air filled layer of thickness $t_{a}$, between the two high-index terminated DBR stacks of the cavity gives fundamental mode frequencies

\begin{footnotesize}
\begin{eqnarray}
\nu_{m} \approx &&\frac{c}{ 2 \pi \left[ t_{a} + \left( n_{d} - 1 \right) t_{d} \right] } \\
&&\times \left( \pi m - (-1)^{m} \arcsin \left( \frac{ n_{d} - 1}{n_{d} + 1} \sin \left( \frac{m \pi \left( t_{a} - n_{d} t_{d} \right) }{t_{a} + n_{d} t_{d}} \right) \right) \right). \nonumber \ \
\label{eqn:resfreq}
\end{eqnarray}
\end{footnotesize}

\begin{figure}[htbp]
	\includegraphics[width=0.46\textwidth]{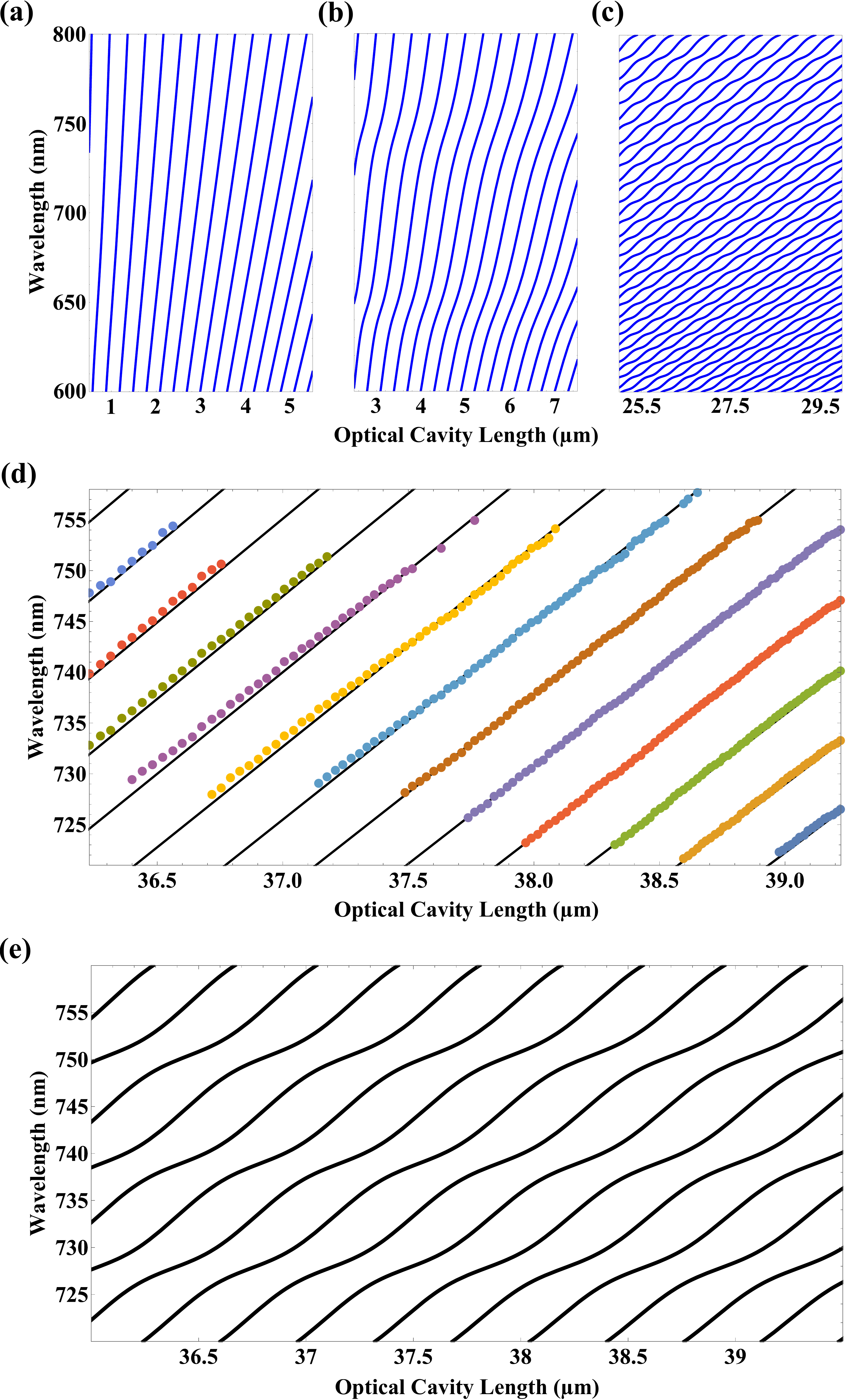}
		\caption{Modification of cavity modes as a function of diamond layer thickness: (a)-(c) Simulation of resonance wavelengths for $t_{d} = (0.2, 1, 10) \, \mu\text{m}$ with an air layer between $t_{a} = 0 - 5 \, \mu\text{m}$ using equation (\ref{eqn:resfreq}). (d) Resonance wavelengths measured (dots) and theory (black curves) of the $0.2 \, \mu\text{m}$ diamond membrane for $L = 36 - 40 \, \mu\text{m}$ cavities. (e) Theoretical distribution of the fundamental mode frequencies of a $10 \, \mu\text{m}$ thick membrane for similar cavity lengths calculated using equation (\ref{eqn:resfreq}).}
	\label{fig:dispersion}
\end{figure}

\begin{figure}[htbp]
	\includegraphics[width=0.46\textwidth]{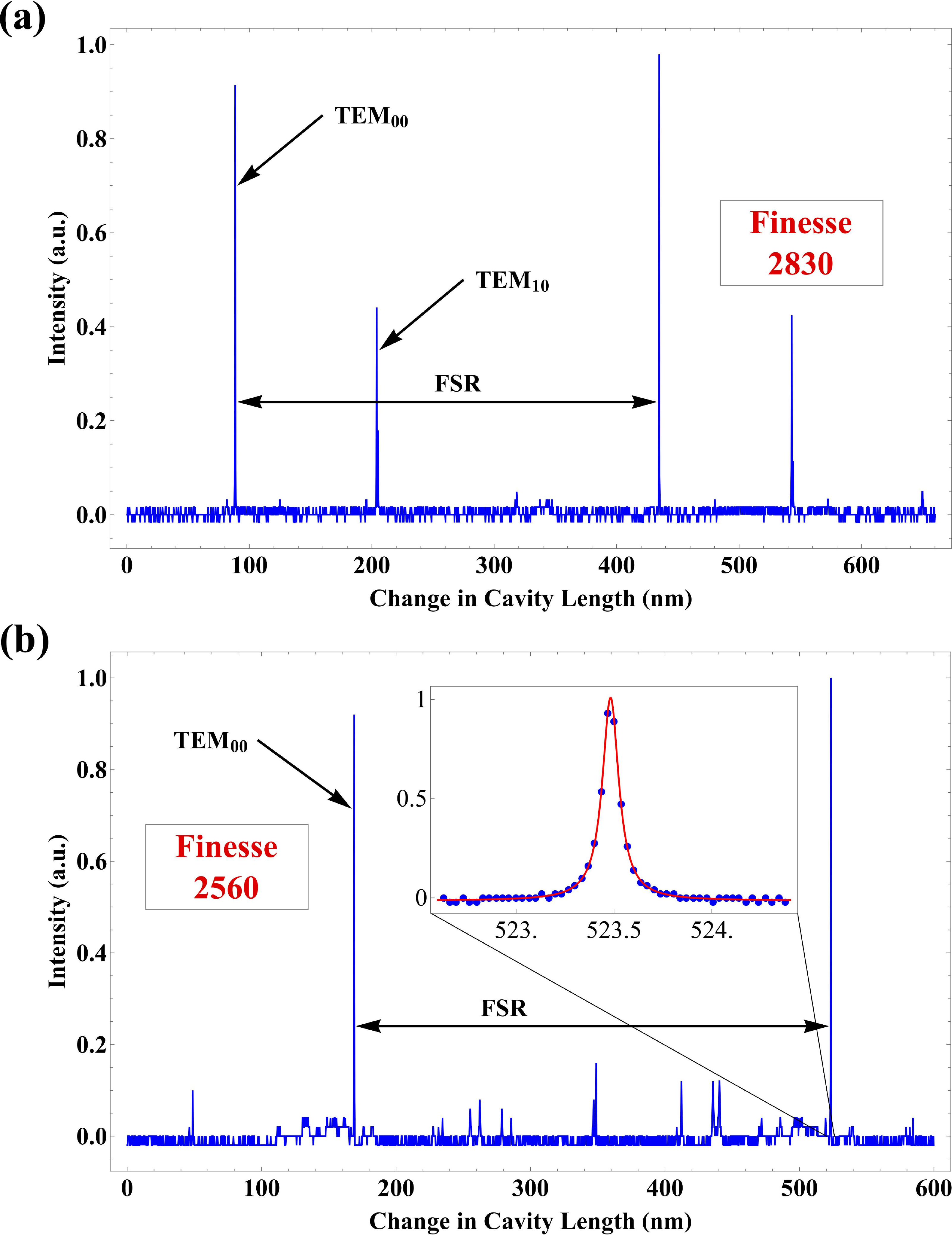}
		\caption{Extracting the Finesse of the setups: Measured cavity transmission as a function of the cavity length for an empty cavity (a) and with a $\approx 200 \, \text{nm}$ thick diamond membrane inside the cavity (b). We normalized the intensities to the maximum transmitted intensity in the TEM$_{00}$ mode respectively. The corresponding Finesse values are $\mathcal{F}_{e} = 2830$ for the empty cavity and $\mathcal{F}_{d} = 2560$ for the diamond membrane cavity system.}
	\label{fig:finesse}
\end{figure}

The effect of the diamond membrane on the modal frequencies increases for thicker membranes, as diamond-like modes (shallow slope) are clearly visible leading to avoided crossings between neighboring modes. This behavior is visualized in the figures \ref{fig:dispersion}(a),(b) and (c) for diamond membranes of different thickness $t_{d} = 200 \, \text{nm}$, $1 \, \mu\text{m}$ and $10 \, \mu\text{m}$, $n_{d} = 2.425$ and the smallest accessible air layer $t_{a} = 0 - 5 \, \mu\text{m}$ respectively. The measured data for our $\approx \, 200 \, \text{nm}$ thick membrane is shown in figure \ref{fig:dispersion}(d) for a cavity length between $36 \, \mu\text{m}$ and $40 \, \mu\text{m}$. 
It matches well with the theoretical simulations (black curves). We use a cubic correction, which is extracted and calibrated from empty resonator measurements for the measured data taking into account the piezo hysteresis when tuning the cavity length. When considering a cavity of the same length but instead including a much thicker membrane (figure \ref{fig:dispersion}(e)), the effect of the diamond in the spectrum is considerable. When the diamond is thin (figure \ref{fig:dispersion}(a),(b)) the wavelength depends linearly on the length $L$. \\
Next we investigate the influence of the membrane on the resonator Finesse by measuring the transmission of a narrowband laser (sub $100 \, \text{kHz}$) at $713 \, \text{nm}$ for both, the empty cavity and the cavity with membrane, while continuously scanning the cavity length over several $\mu\text{m}$. We determine the cavity Finesse $\mathcal{F}$ as the ratio of the free spectral range (FSR) and the full width at half maximum (FWHM) of the resonances as a function of the length of the cavity:
\begin{equation}
\mathcal{F} = \frac{\text{FSR}}{\text{FWHM}}.
\label{eqn:finesse}
\end{equation}
We measure a Finesse of up to $\approx 2800$ for the empty cavity at several different spots on the plane mirror. Figure \ref{fig:finesse}(a) shows the transmitted intensity versus variation of the cavity length, normalized for the maximum transmitted intensity in the TEM$_{00}$ mode. We additionally detect the first order transversal $\text{TEM}_{10/01}$ modes with weaker intensity, due to coupling and alignment mismatch. Lorentzian fitting of the resonance peaks of the fundamental modes yields a Finesse value of $\mathcal{F}_{e} = 2830 \, \pm \, 340$, according to equation (\ref{eqn:finesse}). \\
Inserting the diamond membrane in the cavity does not result in a significant decrease of the transmitted intensity. We still observe Finesse values up to $\approx 2600$ for three different measurement spots on the membrane. Figure \ref{fig:finesse}(b) shows the normalized cavity transmission for the diamond membrane cavity system with a Finesse of $\mathcal{F}_{d} = 2560 \, \pm \, 160$. \\  
The measured Finesse together with the results of the white light spectroscopy allows us to determine the remaining parameters for the diamond membrane cavity system. We obtain the free spectral range directly from the white light spectra (figure \ref{fig:dispersion}(d)): $\Delta \nu_{FSR} = 3.81 \, \pm \, 0.01 \, \text{THz}$ corresponding to an overall cavity length of $L \approx 39 \, \mu\text{m}$. Using $\mathcal{F}_{d} = 2560$ one obtains a cavity linewidth of $\delta\nu = 1.49 \, \pm \, 0.09 \, \text{GHz}$ from equation (\ref{eqn:finesse}), for the diamond cavity system this is below the resolution of our grating spectrometer. Operating at an intermediate cavity length enables us to estimate the coupled SiV$^{-}$-cavity parameters without the risk of damaging the diamond membrane. Additionally the high cavity quality factor of $Q = \lambda/\delta\lambda = \nu/\delta\nu = 2.8 \, \times \, 10^{5}$ (at a wavelength of $713 \, \text{nm}$) improves the photon coherence. \\
In summary we have demonstrated that the insertion of a thin diamond membrane into the cavity does not introduce significant scattering losses. Future perspective arise from integrating SiV$^{-}$ centers in single crystal diamond membranes with ideal optical and spin properties.

\section{Cavity-enhanced absorption measurements to extract the absorption cross section of SiV$^{-}$ center}

\begin{figure}[htbp]
	\includegraphics[width=0.46\textwidth]{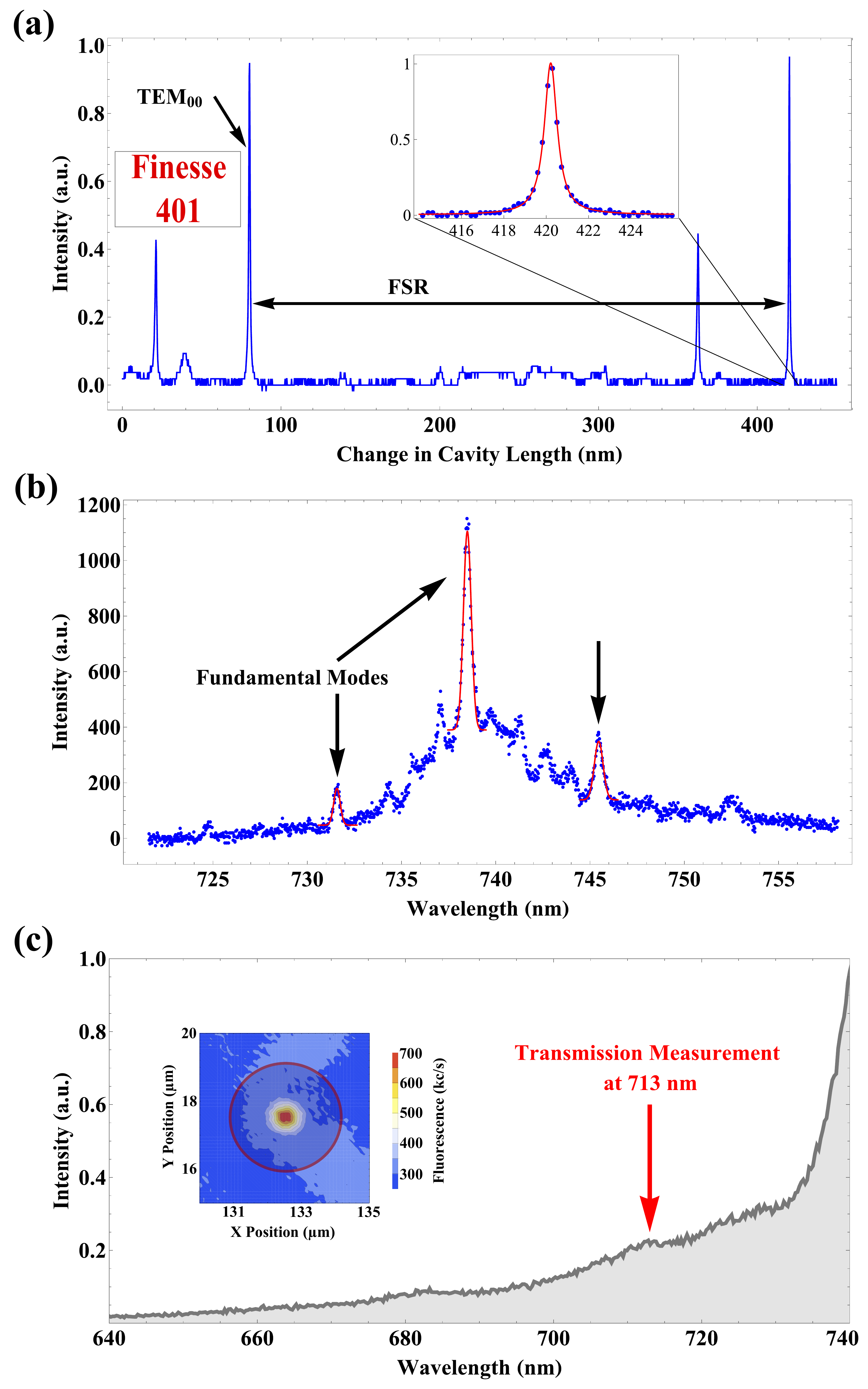}
		\caption{(a) Cavity transmission measurement of the $713 \, \text{nm}$ laser at the position of the SiV$^{-}$ center ensemble resulting in a Finesse of $\mathcal{F}_{SiV} = 401$. (b) PL spectrum of the ZPL of the SiV$^{-}$ ensemble coupled to the cavity mode of a $L \approx 39 \, \mu\text{m}$ cavity. Gaussian fits of the resonance peaks of the fundamental modes (red curves) yielding a linewidth of $264 \, \text{GHz}$. Length jitter and absorption of the defect centers lead to broadening of the resonances. Coupling to higher order transverse modes leads to background and imperfect filtering of the cavity. (c) Mirrored PL spectrum of the SiV$^{-}$ ensemble to illustrate expected absorption of the defect center. The inset shows a 2D confocal image of the SiV$^{-}$ ensemble which has a lateral size (Gaussian FWHM) of $0.68 \, \mu\text{m}$ ($0.64 \, \mu\text{m}$). The red circle indicates the cavity mode waist for comparison.}
	\label{fig:absorption}
\end{figure}

Next, we keep the length of the cavity fixed while laterally moving the flat mirror to overlap the position of the SiV$^{-}$ ensemble with the cavity mode. We measure the Finesse on the SiV$^{-}$ spot from the transmission of the $713 \, \text{nm}$ laser (see figure \ref{fig:absorption}(a)). The resulting value $\mathcal{F}_{SiV} = 401 \, \pm \, 29$ is significantly reduced compared to the measurements of the empty cavity and on the diamond membrane right next to the SiV$^{-}$ ensemble, indicating an additional loss channel introduced by the absorption of the coupled SiV$^{-}$ centers. Scattering losses introduced by the diamond membrane are independently extracted to be negligible by introducing a diamond membrane without SiV$^{-}$ centers into the cavity. From confocal measurements we can exclude any fluorescing defect centers in that region. This is not possible, e.g. when working with nanodiamonds. \\
To confirm the positioning of the SiV$^{-}$ center ensemble in the field maximum of the cavity mode, we study the cavity emission while exciting the SiV$^{-}$ centers with the $532 \, \text{nm}$ laser. We observe coupling of the emitters ZPL to the cavity mode visible as several cavity resonances in the PL spectrum (figure \ref{fig:absorption}(b)). To calculate the spectral enhancement of the SiV$^{-}$ ensemble emission into the resonator mode, we compare the free space emission (figure \ref{fig:confocal}(c)), with the cavity emission (figure \ref{fig:absorption}(b)). We normalize the absolute intensities of the two spectra regarding excitation power and acquisition time leading to a correction factor of $0.42$ for the cavity coupled spectrum and take into account the ten times higher resolution of the grating spectrometer for the cavity measurement. The resulting spectral enhancement factor originating from cavity funneling at the center wavelength of the SiV$^{-}$ ensemble (at $\lambda = 738. 49 \, \text{nm}$) yields $\gtrsim 1.9$.
Gaussian fits of the resonance peaks (red curves in figure \ref{fig:absorption}(b)) yield a linewidth of $264 \, \text{GHz}$ for the fundamental modes. Broadening compared to the natural cavity linewidth calculated in the previous section results from length jitter during the acquisition time of the PL spectrum and extinction introduced by the SiV$^{-}$ centers. In the absence of any cavity jitter, for example for an actively stabilized cavity, we expect an increase of the enhancement factor by more than one order of magnitude. Coupling to higher order transverse modes leads to a background in the emission spectrum. This coupling mainly originates from the dimensions of our SiV$^{-}$ ensemble, which is larger than the resolution limit of our confocal microscope and can therefore not be described as a point source. The SiV$^{-}$ ensemble size together with the cavity mode waist is visualized in the inset of figure \ref{fig:absorption}(c). \\
In order to extract the absorption cross section of a single SiV$^{-}$ center, we estimate the absoption spectrum by mirroring the PL emission at the center ZPL frequency (figure \ref{fig:absorption}(c)) as reported in \cite{haeussler2017photoluminescence}. Absorption of the excitation laser leads to a reduction of the Finesse. The Finesse is evaluated by considering the transmission and losses of the two mirrors, the scattering losses introduced by the diamond membrane and the SiV$^{-}$ ensemble respectively: 
\begin{equation}
\mathcal{F} = \frac{2 \pi}{T_{1} + T_{2} + 2L_{\text{mirror}} + 2L_{\text{membrane}} + 2L_{\text{SiV}}},
\label{eqn:finesse2}
\end{equation}  
where $T_{1}$ and $T_{2}$ are the individually verified transmission values of the two mirrors at a given wavelength, and $L_{\text{mirror}}$, $L_{\text{membrane}}$ and $L_{\text{SiV}}$ are the loss factors of the mirrors, the membrane and the defect centers. We independently determine the individual loss factors by using the measured Finesse values for the empty resonator, the membrane-cavity system and the cavity with the coupled SiV$^{-}$ ensemble of the previous section. At a wavelength of $713 \, \text{nm}$ $T_{1} = T_{2} = 0.0005$ yielding
\begin{eqnarray}
L_{\text{mirror}} &&= \frac{1}{2} \left(\frac{2 \pi}{\mathcal{F}_{e}} - T_{1} - T_{2} \right) \nonumber \\
&&\cong (610 \pm 151) \, \text{ppm},
\label{eqn:lossesmirror}
\end{eqnarray}  
\begin{eqnarray}
L_{\text{membrane}} &&= \frac{1}{2} \left(\frac{2 \pi}{\mathcal{F}_{d}} - T_{1} - T_{2} - 2L_{\text{mirror}} \right) \nonumber \\
&&\cong (117 \pm 183) \, \text{ppm},
\label{eqn:lossesmembrane}
\end{eqnarray}   
and
\begin{eqnarray}
L_{\text{SiV}} &&= \frac{1}{2} \left(\frac{2 \pi}{\mathcal{F}_{SiV}} - T_{1} - T_{2} - 2L_{\text{mirror}} - 2L_{\text{membrane}} \right) \nonumber \\
&&\cong (6607 \pm 618) \, \text{ppm}.
\label{eqn:lossesSiV}
\end{eqnarray}  
From the losses introduced by the SiV$^{-}$ ensemble and a focal spot area $A = \pi \omega_{0}^{2}/4$, corresponding to the resonator waist of the half-symmetric cavity $\omega_{0} \approx \sqrt{\lambda/\pi} (LROC-L^{2})^{1/4} = 3.18 \, \mu\text{m}$ \cite{hunger2010afiber}, we estimate the absorption cross section of the SiV$^{-}$ ensemble to be
\begin{eqnarray}
\sigma_{\text{abs, ensemble}}(713) &&= L_{\text{SiV}} \times A \nonumber \\
&&= 5.25 \, \cdot \, 10^{-10} \, \text{cm}^{2}.
\label{eqn:absorptioncrosssectionens}
\end{eqnarray}  
The biggest uncertainty for extracting the absorption cross section of a single SiV$^{-}$ center arises from the uncertainty in estimating the number of emitters in the SiV$^{-}$ ensemble. We can roughly estimate the number of SiV$^{-}$ centers from the intensity of the spot in the confocal microscope image ($\approx 400 \, \text{kc/s}$) and the $g^{2}(0)$ value. Comparing the count rate with individual SiV$^{-}$ centers measured in our setup with a $\text{NA} = 0.95$ microscope objective (similar to measurements performed in \cite{rogers2014multiple}) we estimate the number of SiV$^{-}$ centers in the ensemble to be between $N = 100$ and $N = 1000$. In our assumptions the single SiV$^{-}$ center emission is averaged over all possible dipole orientations and we take into account the lower collection efficiency of our $\text{NA} = 0.7$ microscope objective resulting in a reduction of the count rate by factor $0.52$. We further consider near-field optic effects and waveguiding originating from the thickness of the diamond membrane close to the emitter wavelength, which can significantly alter the count rates as simulations in reference \cite{neyts1998simulation} show. The point spread function (PSF) of the ensemble is significantly larger than the lateral dimensions of a single SiV$^{-}$ center and is larger than the spatial resolution limit of our microscope. Additionally, we performed second-order correlation measurements with off-resonant excitation ($532 \, \text{nm}$) yielding $g^{(2)}(0) \cong 1$, confirming that there is a large number of SiV$^{-}$ centers in the ensemble. \\
Assuming, that the absorption scales linearly for $N$ independent absorbers, the absorption coefficient of a single SiV$^{-}$ center is accordingly
\begin{eqnarray}
Abs_{\text{SiV}}(713) &&= \frac{L_{\text{SiV}}}{N} \nonumber \\
&&= (3.6 \, \pm \, 3.0) \, \cdot \, 10^{-5}.
\label{eqn:absorptioncoefficient}
\end{eqnarray}  
This leads to a single SiV$^{-}$ absorption cross section at $713 \, \text{nm}$ of
\begin{eqnarray}
\sigma_{\text{abs, SiV}}(713) &&= Abs_{\text{SiV}} \times A \nonumber \\
&&= (2.9 \, \pm \, 2.0) \, \cdot \, 10^{-12} \, \text{cm}^{2}.
\label{eqn:absorptioncrosssection}
\end{eqnarray}  
In reference \cite{neu2012photophysics}, absorption cross sections of $\sigma_{\text{abs, SiV}} = (1.4 \, - \, 4.2) \, \cdot \, 10^{-14}  \, \text{cm}^{2}$ were measured for a single SiV$^{-}$ center with off-resonant excitation at $532 \, \text{nm}$. \\
To theoretically estimate the value of the absorption cross section we suppose a two-level system where the absorption cross section is given by 
\begin{equation}
\sigma_{\text{abs}} = \frac{3 \lambda^{2}}{2\pi} \frac{\gamma}{\gamma^{*}}
\label{eqn:theoabsorptioncrosssection1}
\end{equation}  
with spectral broadening taken into account by the ratio $\gamma/\gamma^{*}$ and further consider dependencies on the Debye-Waller factor ($\alpha_{\text{DW}}$) as reported in reference \cite{wrigge2008efficient} and on a factor describing the fraction of non-radiative decay ($\alpha_{\text{NR}}$). The Debye-Waller factor of the SiV$^{-}$ center is $\alpha_{\text{DW}} \sim 0.7$ \cite{rogers2014multiple}, while $\alpha_{\text{NR}}$ has not been exactly determined so far. Additionally we take into account the detuning of the excitation laser by $\Delta \sim 28 \, \text{nm}$ from the ZPL, reducing the absorption by a factor of $1/\alpha_{\text{OR}} \approx 6$ as extracted from the references \cite{kern2017optical, haeussler2017photoluminescence}. We conclude with
\begin{eqnarray}
\sigma_{\text{abs, SiV}}(713) &&= \alpha_{\text{NR}} \times \alpha_{\text{DW}} \times \alpha_{\text{OR}} \times \frac{3 \lambda^{2}}{2\pi} \frac{\gamma}{\gamma^{*}} \nonumber \\
&&\approx \alpha_{\text{NR}} \times 9.1 \, \cdot \, 10^{-15} \, \text{cm}^{2}
\label{eqn:theoabsorptioncrosssection2}
\end{eqnarray}  
for a single SiV$^{-}$ center with $\gamma / 2 \pi = 94 \, \text{MHz}$ and $\gamma^{*} / 2 \pi = 2.92 \, \text{THz}$. The room temperature dephasing rate $\gamma^{*} / 2 \pi$ for a single SiV$^{-}$ center was measured seperatly in bulk diamond and we believe that we are able to observe these parameters in future membrane samples with a combination of ion implantation and surface termination methods, as it was recently shown for nanodiamonds in reference \cite{rogers2019single}. For $\alpha_{\text{NR}} \sim 0.12$ this is in very good accordance with recent derivations of the emission and absorption cross section from PL spectra and McCumber theory in reference \cite{kern2017optical}. Following their calculations we find $\sigma_{\text{abs, SiV}}(713) \approx 1.1 \, \cdot \, 10^{-15} \, \text{cm}^{2}$ at $\lambda = 713 \, \text{nm}$ and about a factor of $6$ higher at resonance: $\sigma_{\text{abs, SiV}}(738) \approx 6.1 \, \cdot \, 10^{-15} \, \text{cm}^{2}$. Usually the effect of non-radiative relaxation should be negligible. Fast non-radiative decay should even lead to an increased absorption. In case of a long-lived shelving state however the absorption can be significantly reduced, resulting in $\alpha_{\text{NR}} < 1$. Such a dark or shelving-state was proposed by ab initio studies in reference \cite{gali2013abinitio} and by a rate analysis in reference \cite{benedikter2017cavity}. A by a factor of $\sim 10$ reduced quantum yield due to non-radiative decay was also reported in reference \cite{rogers2014multiple} which is consistent with our approximated value of $\alpha_{\text{NR}} = 0.12$. \\
Experimentally we observe a value for the absorption cross section that significantly deviates from theory predictions by at least two orders of magnitude. A similar tendency was also reported in reference \cite{neu2012photophysics}. A higher precision in the determination of $\sigma_{\text{abs, SiV}}$ can be achieved in future measurements with single SiV$^{-}$ centers.

\section{Discussion}

In our work we investigate an ensemble of SiV$^{-}$ centers in a thin diamond membrane coupled to the mode of a fiber-based optical resonator as a potential light-matter interface for applications such as quantum repeater nodes. We explore the dispersion relation and demonstrate a linear dispersion relation, as for an empty resonator and the Finesse is unaffected by the presence of the membrane. Our findings enable experiments with high-Finesse resonators and with very small resonator lengths below $1 \, \mu\text{m}$ and limited only by the geometry of the cavity mirrors. \\
Extracting from our parameters, one finds coupling strengths of $g_{0} / 2 \pi \sim 18 \, \text{GHz}$ for single SiV$^{-}$ centers may be possible. Together with fast field decay rates of $\kappa / 2 \pi \sim 50 \, \text{GHz}$ an efficient spin-photon interface can be realized. We extract the absorption cross section from the reduction in Finesse to be $(2.9 \, \pm \, 2.0) \, \cdot \, 10^{-12} \, \text{cm}^{2}$. The largest uncertainty in our measurement arises from the uncertainty in the estimated number of emitter. In future experiments the absorption cross section can be determined with high precision when the experiment is repeated with single SiV$^{-}$ center. \\
One possible interpretation for the much larger experimentally estimated absorption cross section compared to theoretical predictions could be a (partly) collective absorption \cite{higgins2014superabsorption}. Although at first glance this might be surprising since the inhomogeneous ensemble linewidth of $\gamma_{\text{ensemble}}^{*} / 2 \pi \sim 4.86 \, \text{THz}$ is much larger than the lifetime-limited linewidth of $\gamma_{\text{SiV}} / 2 \pi = 94 \, \text{MHz}$, the parameters of the coupled SiV-cavity system render superabsorption possible. As described in reference \cite{auffeves2011few} the coupling of light to the collective dipole is given by $\Gamma = 4 g_{0}^{2}/\kappa$ yielding $\Gamma / 2 \pi \sim 4.6 \, \text{GHz}$ for our settings and assuming a weak pumping rate. We characterize the collective behaviour by $\Gamma/\gamma \sim 49 > 1$ indicating a significant collective absorption. However, dephasing and large inhomogeneous ensemble linewidth causes the collective spin to individualize proportional to $\Gamma/\gamma^{*} \sim 10^{-3} < 1$. In case of a large ensemble we still expect partly collective absorption despite strong dephasing by a SiV$^{-}$ subensemble. The total ensemble absorption coefficient is then approximated by $Abs_{\text{SiV}} = L_{\text{SiV}}/(N+M^{2})$ where $M$ is the number of SiV$^{-}$ center acting as single collective superabsorber. Comparing the experimental value for $L_{\text{SiV}} = 6.6 \cdot 10^{-3}$ with the theoretical absorption coefficient for a single SiV$^{-}$ center $Abs_{\text{SiV, theo}} = 1.38 \cdot 10^{-8}$ (reference \cite{kern2017optical}) puts an upper limit for partial superabsorption for an ensemble size $< 4.8 \cdot 10^{5}$. In particular, for non-uniformly distributed transition frequencies the onset of superabsorption becomes feasible. Temperature dependent measurements of the absorption cross section could unfold the effect of superabsorption. Cooling the sample to $\sim 5 \, \text{K}$ could decrease the dephasing of $\gamma^{*} < \Gamma$ with the potential for a large increase in superabsorption. \\
In future experiments our introduced platform can be utilized towards the creation of indistinguishable single photons at room temperature. Optimizing the cavity parameters for cavity funneling yields a degree of indistinguishability $> \, 0.8$ at rates up to tens of MHz (see Supplement Material for details).

\section*{Acknowledgements}

Experiments performed for this work were operated using the Qudi software suite \cite{binder2017qudi}. We thank Rodolfo Previdi for his help and expertise during the fabrication of the diamond membrane. We further thank Mathias Metsch and Petr Siyushev for experimental support. SH and AK acknowledge support of IQst. AD and AK acknowledge support of the Carl-Zeiss Foundation. JT acknowledge funding from the Australian Research Council Centre of Excellence in Engineered Quantum Systems (Grant No CE170100009). IA acknowledges the generous support provided by the Alexander von Humboldt Foundation and the Australian Research council (DP180100077). DH and AK acknowledge support of BMBF in project Q.Link.X. AK acknowledges support of the Wissenschaftler-Rückkehrprogramm GSO/CZS and DFG.

\section*{Author Contributions}

SH and AK conceived the experiments, performed the measurements and evaluated the data. JB and DH fabricated the fiber resonators. KB, BR and IA performed the generation of the diamond membranes. The manuscript was written by SH and AK and all authors discussed the results and contributed to the manuscript.

\newpage

\bibliography{Diamond_Photonics_Platform}

\end{document}